\documentclass[letterpaper,twocolumn,10pt]{article}
\usepackage{usenix,epsfig,bytefield}
\usepackage{ifthen}
\usepackage{xcolor}
\usepackage{booktabs}
\usepackage{color}
\usepackage{colortbl}

\usepackage{float}                           

\newcommand{\COLOR}{yes}
\ifthenelse{\equal{\COLOR}{yes}}{%
  \usepackage[colorlinks,citecolor=blue,breaklinks]{hyperref}%
}{%
  \usepackage[pdfborder={0 0 0},breaklinks]{hyperref}%
}
\usepackage{url}

\usepackage{breakurl}

\begin{document}

\newcommand{\System}{LayerZero}
\newcommand{\SystemEndpoint}{\System{} Endpoint}

\newcommand{\showComments}{yes}
\newcommand{\note}[2]{
    \ifthenelse{\equal{\showComments}{yes}}{\textcolor{#1}{#2}}{}
}
\newcommand{\nt}[1]{\note{blue}{Note: #1}}

\title{\Large \bf \System{}: Trustless Omnichain Interoperability Protocol}

\author{
 {\normalfont Ryan Zarick} \\ \\
\and
 {\normalfont Bryan Pellegrino} \\ \\  {\normalfont May 26, 2021}
\and
  {\normalfont Caleb Banister} \\ \\ 
}
\maketitle

\begin{abstract}
The proliferation of blockchains has given developers a variety of platforms on which to
run their smart contracts based on application features and requirements for throughput, security, and cost.
However, a consequence of this freedom is severe fragmentation;
Each chain is isolated, forcing users to silo their liquidity and limiting options to move
liquidity and state between walled ecosystems.

This paper presents \System{}, the first \emph{trustless} omnichain interoperability protocol, which
provides a powerful, low level communication primitive upon which a diverse set of cross-chain
applications can be built. Using this new primitive, developers can implement seamless inter-chain applications like a
cross-chain DEX or multi-chain yield aggregator without having to rely on a trusted custodian or
intermediate transactions.
Simply put, \System{} is the first system to trustlessly enable direct transactions across all chains.
Allowing transactions to flow freely between chains provides opportunities
for users to consolidate fragmented pockets of liquidity while also making full use of applications on
separate chains. With \System{}, we provide the network fabric underlying the fully-connected omnichain ecosystem of
the future.
\end{abstract}

\section{Introduction}
\label{sec:intro}

At the core of the blockchain concept are the three pillars of decentralization, transparency, and immutability.
No single entity controls the blockchain, and actions on the blockchain are verifiable and irreversible.
These pillars create a foundation upon which an entity can act without trusting any other entity.
This trust guarantee is one reason why, for example, cryptocurrencies are enticing compared to fiat currency.

If all users and all applications coexisted in one unified blockchain, then this paper would conclude here.
However, the utility of the blockchain has led to a proliferation of diverse applications, with unique intricacies and requirements.
The demand for a diverse set of functionalities spurned the growth of specialized chains.
Each of these chains has fostered immense growth in applications within its own ecosystem, but the \emph{isolation} between these ecosystems
has emerged as a key limit to adoption. Users and developers are forced to split time, resources, and liquidity between separate chains.
A natural consequence of the sheer number of so-called Layer 1 blockchains (as many as 109 at the time of writing~\cite{blockchain-comparison})
is the need to extend the above-mentioned three pillars to envelope interactions across multiple chains simultaneously.
One example of an in-demand interaction between chains is the transfer of tokens, which we discuss later in this section.

\begin{figure}
    \centering
    \includegraphics[width=\linewidth]{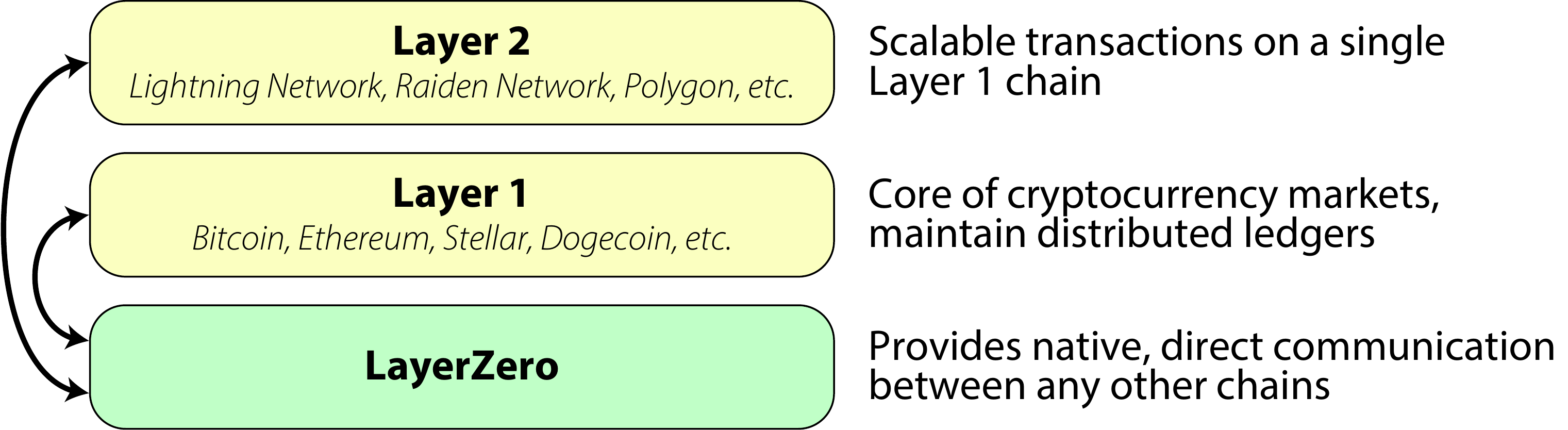}
    \caption{\System{} enables cross-chain transactions.}
    \label{fig:layers}
\end{figure}

In blockchain parlance, the unit of work is a transaction, immutable and irrevocable.
Transactions, collated into blocks, form the basis of security in a blockchain system.
However, transactions have always been a single-chain concept; as described below,
interacting across chains has traditionally required a third-party mechanism outside of the normal blockchain cryptosystem.
In contrast, this paper describes the first messaging protocol upon which native cross-chain transactions are possible: \System{}.

To illustrate the powerful communication primitive \System{} provides, we look at the example of transferring tokens from one chain to another.
Currently, to convert between tokens of two chains, a user must leverage either a centralized exchange
or a cross-chain decentralized exchange (DEX) (also known as a cross-chain bridge),
both of which require a compromise.
In the case of a centralized exchange, e.g., Binance.com~\cite{binanceexchange}, the user must trust the exchange that is tracking deposits and funding withdrawals.
This trust relationship is contrary to the fundamental trustlessness of blockchain consensus
and lacks the security of an on-chain automated system.
Using a DEX, such as AnySwap~\cite{anyswap} or THORChain~\cite{thorchain}, alleviates the trust problem by conducting the transfer on-chain, but existing DEX implementations involve converting the user token 
into a protocol-specific token that transits their intermediate consensus layer to achieve transaction consensus.
This intermediate consensus layer, though usually implemented in a secure manner, does require the user to trust a
side chain to facilitate the token transfer. As we show in this paper, this additional overhead is unnecessary.
Despite heavy user demand, no solution has emerged that is both efficient, direct, while still preserving the core reason for
using blockchains in the first place: trustlessness.
Taking a step back, \System{}'s direct cross-chain transactions gives developers the tools to build just that.

It is important to note that \System{} and the exchanges described above operate at two different levels of the implementation stack.
\System{} is a communication primitive that enables diverse omnichain applications, whereas an exchange is one example of an
application that would benefit from re-implementation on top of \System.
Section~\ref{sec:background} outlines the blockchain technology landscape and explores the exchange example further. 

To properly explain the capabilities of \System{} and its role in the blockchain ecosystem, we first present a formalization of
the fundamental communication primitive required to enable inter-chain transactions, which we term \emph{valid delivery} (Section~\ref{sec:valid-delivery}).
We then describe how \System{} provides this primitive in a trustless manner, thus preserving the security promise of blockchain.
\System{} is the first trustless omnichain interoperability layer, and supports messaging directly between both Layer 1 and Layer 2 chains (Figure~\ref{fig:layers}).

\begin{figure}
    \includegraphics[width=\linewidth]{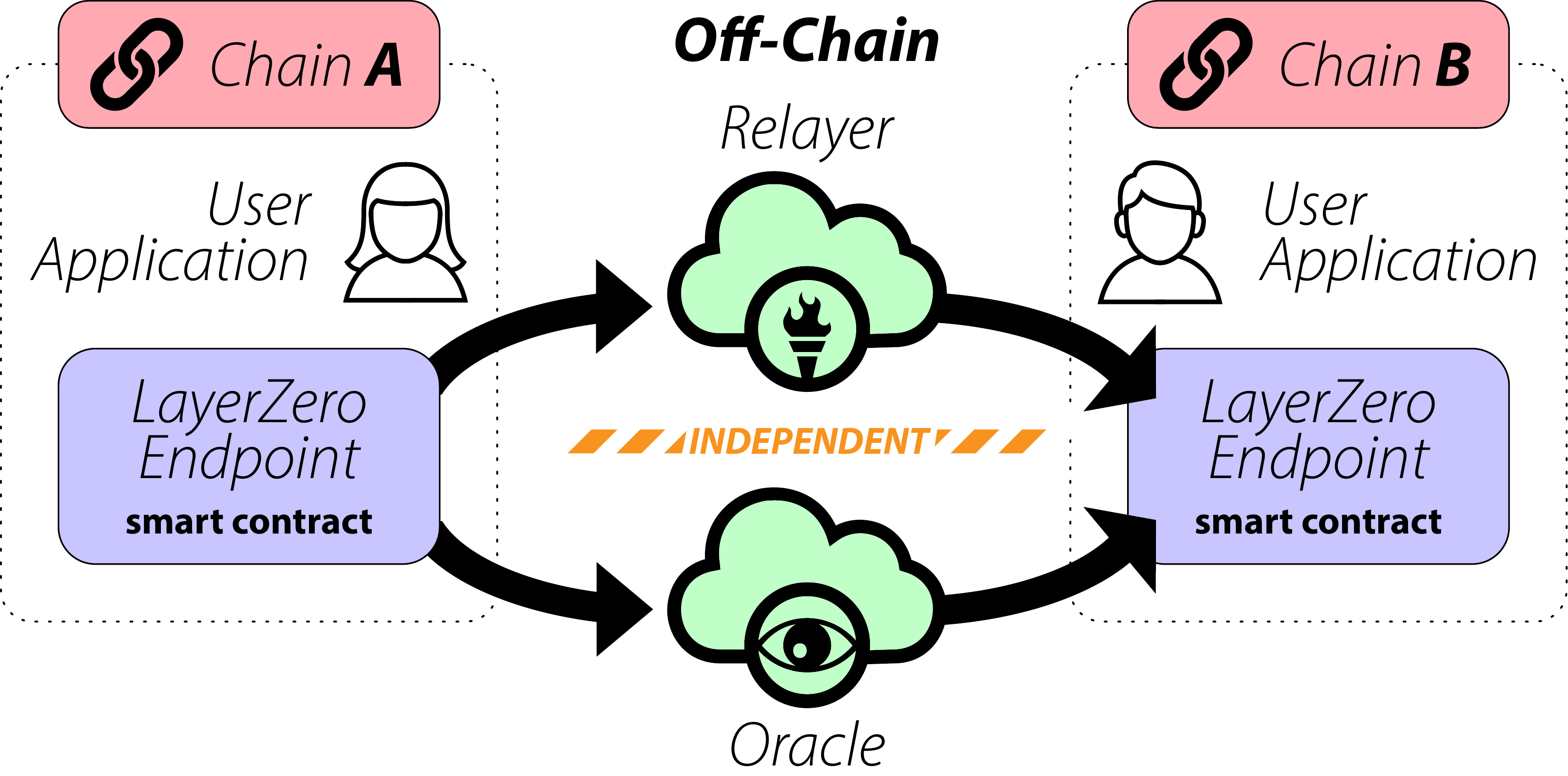}
    \caption{\System{} ensures the validity of cross-chain communication by requiring that two independent entities, the Oracle and Relayer, corroborate the transaction.}
    \label{fig:overview}
\end{figure}

A cross-chain transaction between chains $A$ and $B$ consists of a transaction $t_A$ on $A$, a communication protocol between $A$ and $B$, and a message $m$.
Valid delivery states that $m$ is delivered if and only if $t_A$ is committed and valid. 
The key idea underpinning \System{} is that if two \emph{independent} entities corroborate the validity of a transaction (in this case, $t_A$)
then chain $B$ can be sure that $t_A$ is valid. Figure~\ref{fig:overview} illustrates this at a high level.
Given two entities that do not collude, if
(1) one entity can produce a block header for the block containing $t_A$ on chain $A$,
(2) the other entity can \emph{independently} produce the proof for $t_A$ on that block (transaction proof),
and (3) the header and transaction proof in fact agree, then
the communication protocol can deliver $m$ to the client on chain $B$ with the guarantee that $t_A$ is stably committed on chain $A$.
The \System{} communication protocol, described in Section~\ref{sec:design}, guarantees that the transaction on the recipient chain will be paired with a valid, 
committed transaction on the sender chain  without involving any intermediary chains.
We achieve this by combining two independent entities: an Oracle~\cite{chainlink17} that provides the block header, and a \emph{Relayer}
that provides the proof associated with the aforementioned transaction. 

The interface to \System{} is a lightweight on-chain client, which we call the \SystemEndpoint{}.
One \SystemEndpoint{} exists on each (supported) chain, and any chain with a \SystemEndpoint{}
can conduct cross-chain transactions involving any other chain with a \SystemEndpoint{}.
In essence, this creates a fully-connected network where every node has a direct connection to
every other node. With minor boilerplate code, any blockchain is supported. Section~\ref{sec:casestudy} demonstrates this process through a
case study in implementing \System{} on the Ethereum blockchain.

The ability to perform cross-chain transactions directly with any other chain
on the network opens the opportunity for a class of large-scale applications that were previously infeasible,
such as cross-chain decentralized exchanges, multi-chain yield aggregators, and cross-chain lending.
Section~\ref{sec:apps} examines several such applications in detail. Through \System{}, users can freely move
liquidity between chains, allowing for a single pool of liquidity to take part in multiple decentralized finance (DeFi)
applications across different chains and ecosystems without having to go through third party systems or intermediate tokens.

\section{Background}
\label{sec:background}

To lay the groundwork for \System{}, we review relevant existing systems to illustrate why they fall short of meeting the demands of emerging applications. The discussion culminates in an in-depth explanation of the advantages in building a cross-chain exchange atop \System{}.

\subsection{Related work}
\label{sec:related}

This section builds an understanding of the important players in the cross-chain interaction space,
why they fall short of the ideals of trustless valid delivery, and how \System{} closes that gap.

\textbf{Ethereum~\cite{eth}} is the most popular platform for decentralized finance applications built via smart contracts.
Ethereum extends its underlying blockchain with a Turing-complete programming language that enables a library of decentralized applications to leverage the
powerful security properties of the underlying chain through a developer-friendly abstraction.
However, the low transaction rate of the underlying blockchain, approximately 15--45 transactions per second~\cite{eth-rate}, has proven to be a
serious scalability bottleneck that limits the popularity of the applications built to run directly on the Ethereum blockchain. Because of its programming model and popularity, many inter-chain communication techniques revolve around interfacing third-party chains with Ethereum.
\System{} provides the ability to directly transfer state to and from Ethereum without a middleman, allowing users and applications to leverage the
stability and trustworthiness of the Ethereum chain without the cost and bottlenecks of the solutions described below.

\textbf{Ethereum 2.0~\cite{eth2}} is a set of proposed upgrades to address the scalability, security, and sustainability shortcomings in Ethereum.
Ethereum 2.0 introduces shard chains that distribute load instead of concentrating all transactions on the overloaded Ethereum main chain.
Transitioning from proof-of-work to proof-of-stake both eliminates the possibility of a 51\% attack and reduces the energy per transaction.
These advancements are largely orthogonal to \System{} except that they are sure to boost the popularity of Ethereum, creating even more demand for convenient and cheap inter-chain communication.

\textbf{Polygon~\cite{polygon}}, formerly Matic Network, is a Layer 2 network that addresses the throughput and sovereignty challenges of Ethereum.
Despite being the most popular platform for blockchain development, Ethereum is plagued by low throughput~\cite{galal2019efficient}, making it unsuitable for
certain applications.
Polygon provides application-specific, Ethereum-compatible sidechains that combine the scalability and
independence of separate chains with the community and security of Ethereum.
Specialized or throughput-intensive applications run on the sidechains and periodically consolidate back to the main Ethereum chain.
In contrast, \System{} is a lower-level platform that enables direct inter-chain communication and can be used to facilitate transfers back to the Ethereum chain
without the complexities of the Polygon protocol.

\textbf{Polkadot~\cite{polkadot}} is an early example of the potential of an open cross-chain ecosystem.
In Polkadot, many domain-specific, parallel chains (``parachains'') connect via a common relay chain that enables tokens and data to flow between them.
However, inter-chain communication always crosses this relay chain, thus incurring additional costs.
\System{} provides the same low-level communication platform of Polkadot, without involving the extra transactions necessitated by the on-chain middleman.

\textbf{THORChain~\cite{thorchain}} is a DEX that uses pairwise liquidity pools to transfer tokens between third-party chains.
Each liquidity pool binds a specific third-party currency to a THORChain native token called RUNE, which acts as a common interchange medium.
Without this common medium, all pairs of currencies would need a liquidity pool, meaning that the number of pools would scale as the square of the number of currencies. 
Unfortunately, while RUNE solves this scalability problem, it is a cumbersome overhead in the transaction process that makes a simple operation quite complicated.
This is evident in the complexities of the THORChain transaction algorithm.
\System{} provides direct inter-chain communication without THORChain's inherent scalability bottleneck, cumbersome intermediate currency, or heavyweight protocol.

\begin{figure*}
    \centering
    \includegraphics[width=\linewidth]{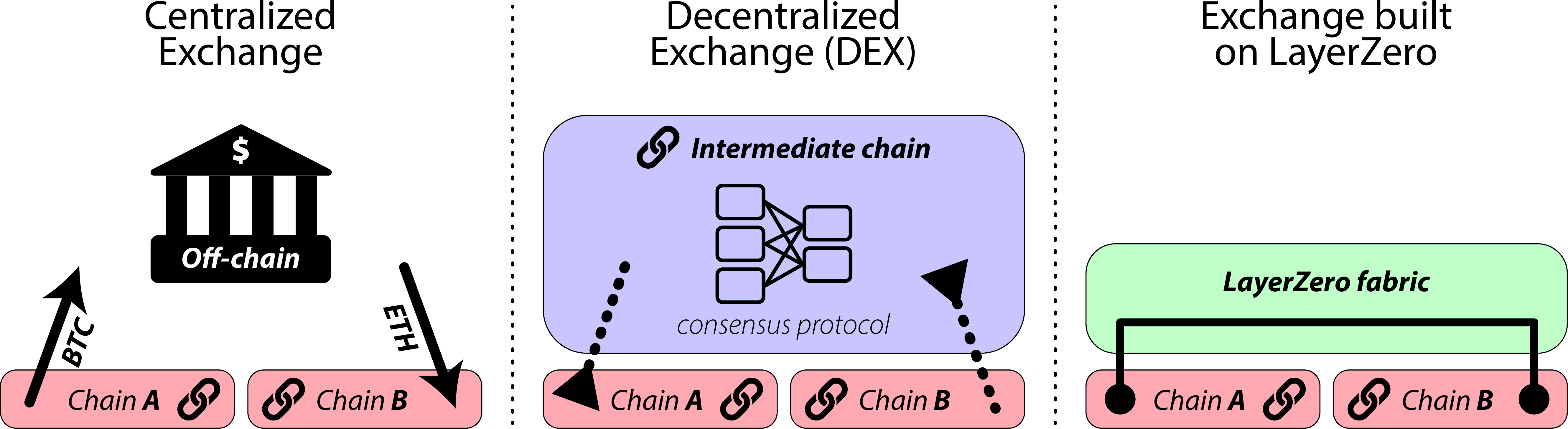}
    \caption{\System{} is a building block for cross-chain applications. This figure visualizes the architectural differences between a centralized exchange, a decentralized exchange, and a cross-chain bridge built using \System{} as its underlying communication primitive.}
    \label{fig:exchanges}
\end{figure*}

\textbf{AnySwap~\cite{anyswap}} is a DEX geared towards easy pairwise token exchanges, similar to THORChain.
AnySwap relies on an intermediate token, ANY, based on Fusion distributed control rights management~\cite{fusion}.
Like with THORChain, the use of the ANY intermediate token introduces undesirable overhead, delay, and additional transfer fees. 


\textbf{Cosmos~\cite{cosmos}} is a blockchain network technology that allows arbitrary messages to be sent between supported chains.
Cosmos includes an Inter-Blockchain Communication (IBC~\cite{cosmos-ibc}) protocol build on Tendermint BFT~\cite{tendermint} to facilitate messaging between
chains built on Cosmos Hub.
Cosmos differs from \System{} in two key ways: (1) IBC runs a full on-chain light node, and (2) IBC only provides direct
communication between fast-finality~\cite{viriyasitavat2019new} chains.
These limitations of IBC, combined with its use of an intermediate chain to facilitate consensus, make it similar to Anyswap, THORChain, or Polkadot,
rather than a general communication layer like \System{}.
Cosmos also provides a DEX with similar properties to Anyswap or THORChain, called the Gravity Bridge~\cite{gravitybridge}.
In contrast to Cosmos and IBC, \System{} provides \emph{trustless} omnichain messaging, and can be extended to run on any chain,
including those which provide probabilistic-finality, such as Ethereum and Bitcoin.

\textbf{Chainlink~\cite{chainlink17, chainlink21}} is a framework for building and connecting to decentralized oracle networks (DONs). While smart contracts are tamperproof, their on-chain nature prevents the basic connectivity crucial to their wider adoption: a smart contract cannot fetch off-chain data that is necessary to the execution of the contract, such as stock prices, IoT device measurements, and outputs from secure off-chain computations. A DON extends a smart contract’s tamperproof property to the data sources and external resources that the contract depends on, without placing trust in any central entity. In a DON, a user’s smart contract makes an on-chain request to a Chainlink interface smart contract, which posts an event to many separate oracle nodes. Each oracle node queries multiple data sources for the requested information, aggregates it to filter erroneous or malicious sources, and optionally performs trust-minimized computations on the data. The oracle nodes respond to the Chainlink interface contract, which performs a second level of aggregation to filter erroneous or malicious oracles. This dual-layer filtering guarantees trust in the final data without requiring trust in any individual oracle or data source. As a result, Chainlink provides a robust information-retrieval network as well as a secure off-chain computation solution that has become widely used across the industry. By leveraging the Chainlink DON framework, the \System{} protocol gains the ability to ensure trustless delivery of messages between disparate chains.

\subsection{\System{} in practice}
\label{sec:exchanges}

Developers can use \System{} to build complex cross-chain applications without sacrificing trustlessness or introducing complex intermediate chains/smart contracts. 
Figure~\ref{fig:exchanges} illustrates the functionality of \System{} in the context of building an exchange.

A centralized exchange, shown on the left, requires users to deposit their tokens with a central trusted authority, which then
keeps track of that deposit off-chain and grants coins on other chains as the user requests them.
Placing trust in this authority defeats the purpose of using blockchain to begin with, which has resulted in the emergence of distributed exchanges.

The center diagram shows, on a high level, how a typical decentralized exchange works---by using a smart contract--governed consensus protocol to facilitate the automatic
minting of coins on chain $B$, DEXs are able to overcome the necessity for a centralized, trusted off-chain middleman.
However, one key limitation is that DEXs involve an intermediate token and intermediate chain, and only mints an intermediary or wrapped token on chain B as opposed
to the actual token the user wants. The user must then exchange the intermediary token (e.g. RUNE) or wrapped token (e.g. ANY) for the desired token in an additional
transaction. This intermediary/wrapped token, second transaction, as well as the intermediate chain are all
unnecessary overheads to what should ideally be a single seamless transaction.
The right side of Figure~\ref{fig:exchanges} shows what an exchange built on \System{} would look like, with chain $A$ able to initiate a single
cross-chain transaction that facilitates the local transaction on chain $A$ and notifies the application on chain $B$ that they can safely grant a token
to the user. In this application, \System{} enables a clean and minimal single-transaction swap that does not include any intermediate tokens.
The actual exchange protocol is handled by smart contracts on either side of the cross-chain transaction, with \System{} delivering
messages between the two. This provides a great deal of flexibility, and follows the end-to-end principle~\cite{saltzer1984end} with the majority
of the high level exchange logic handled by smart contracts on the source and the destination chains.

\section{Valid Delivery}
\label{sec:valid-delivery}

In this section, we describe the fundamental properties of trustless inter-chain communication.
To formally characterize the problem of validating a transaction on a different chain, we define the idea of \emph{valid delivery}.
Valid delivery is a communication primitive that enables cross-chain token transfer by providing the following guarantees:

\begin{enumerate}
    \item Every message $m$ sent over the network is coupled with a transaction $t$ on the sender-side chain.
    \item A message $m$ is delivered to the receiver if and only if the associated transaction $t$ is valid and has been committed on the sender-side chain.
\end{enumerate}

Centralized exchanges guarantee valid delivery, in that the agreement between the client and the exchange is that the client will transfer their token
from one chain to the exchange, and the exchange will, upon receipt of that token, issue some balance (non-cryptocurrency). This non-cryptocurrency balance
can then be withdrawn from any available chain, a convenience which is made possible by broad pools of liquidity maintained by the exchange on each of the supported chains.
The exchange acts as the middleman in this transaction and the user must trust them to uphold their end of the bargain.
However, a malicious or compromised exchange could take tokens from the client, issue balance, and then refuse to allow withdrawal of that balance from another chain,
effectively stealing tokens from the user. Even if the user is willing to trust the exchange, recent years have seen many successful
attempts to hack or compromise cryptocurrency exchanges~\cite{lazarenko2018financial}, so users are better served with a solution that
does not require any trusted middleman.
At a higher level, one of the core tenants of cryptocurrencies is their independence from centralized entities like banks,
so relying on a centralized exchange defeats their purpose.

\begin{figure*}
    \includegraphics[width=\linewidth]{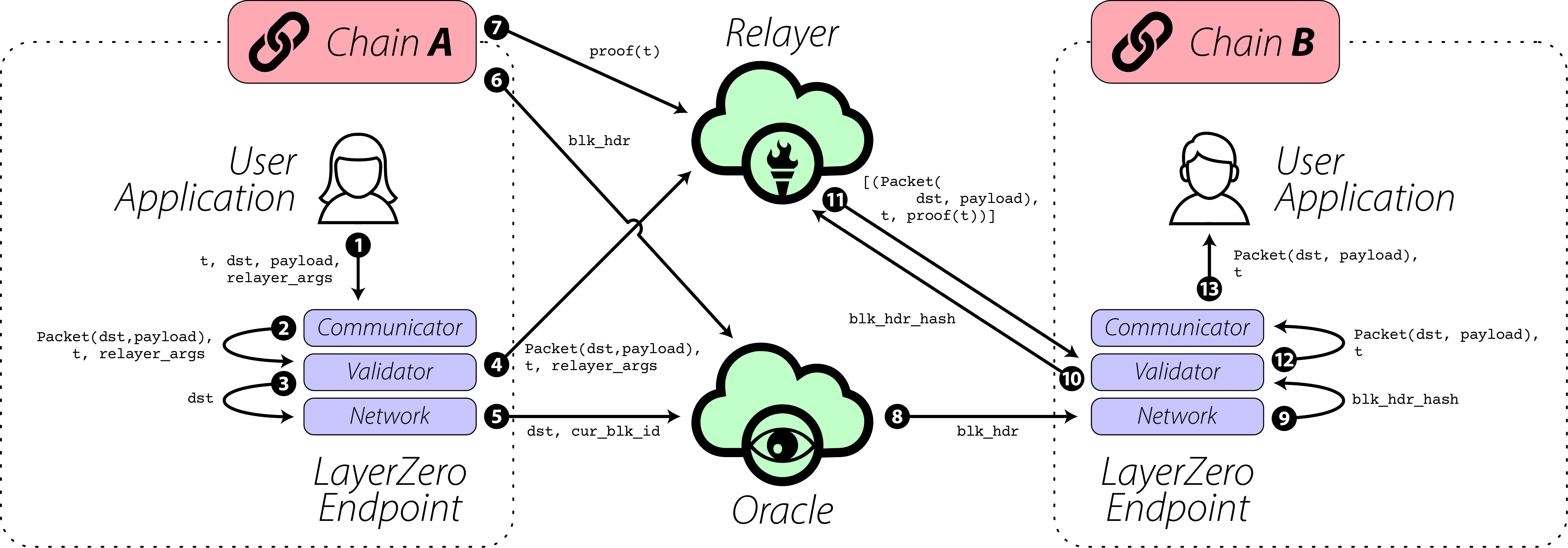}
    \caption{The communication flow in a single \System{} cross-chain transaction.}
    \label{fig:commflow}
\end{figure*}

An alternative to using an centralized exchange is a decentralized exchange such as THORChain~\cite{thorchain} or AnySwap~\cite{anyswap}.
All existing DEX use an intermediary token, such as RUNE in the case of THORChain or ANY in the case of AnySwap, as
the transaction $t$. Because these intermediary tokens are governed by the respective protocols of each DEX, the DEX is able to guarantee valid
delivery, as it is impossible for a malicious user to fake the intermediary token.
Existing DEX solutions are not ideal because they involve two intermediate transactions---one to convert the sender's
token to an intermediary token and one to convert the intermediary token to the desired ``real'' token on the recipient chain.
In addition to this, it is necessary for the user to fully trust the intermediate consensus layer that confirms the transaction
on the source chain and conveys intent to mint the token to the destination chain.
While existing exchanges do enable cross-chain token transfer, they do so at the cost of unnecessary complexity and cost.
The downsides of this are evident in that cross-chain applications have not seen broad adoption.
The ideal solution to the inter-chain transaction problem is one that uses a single one-swap transaction between chains without involving
any trusted middle entity---in other words \emph{trustless} valid delivery. In our work, we implement a \emph{generic messaging
protocol} that provides trustless valid delivery of arbitrary user data, not just tokens.
Distributed exchanges or other DeFi applications would be implemented using our messaging primitive to provide cross-chain
transactions, and the degree of flexibility provided by a low-level messaging protocol enables higher-level applications to implement
a wide range of previously-impossible functionality.

\section{Design}
\label{sec:design}

The core of \System{} is a communication protocol that provides trustless valid delivery. Our protocol
is built on a series of components introduced in Section~\ref{sec:components}.
We discuss the communication flow of the transfer protocol in Section~\ref{sec:protocol},
describe how \System{} is able to achieve valid delivery without involving trusted intermediary services in Section~\ref{sec:tvd}, and
present our novel design for a low-cost smart contract--based light client endpoint in Section~\ref{sec:endpoint}.

\subsection{\System{} components}
\label{sec:components}

\textbf{\SystemEndpoint{}s} are the user-facing interface to \System{}. Each chain in the \System{} network has one \SystemEndpoint{}
implemented as a series of on-chain smart contracts. An Endpoint's purpose is to allow the user to send a message using the \System{} protocol
backend, guaranteeing for valid delivery.

A \SystemEndpoint{} is split into four modules: Communicator, Validator, Network, and Libraries.
The Communicator, Validator, and Network modules comprise the core functionality of the Endpoint (Figure~\ref{fig:commflow}), while each
new chain supported by \System{} is added as an additional Library.
This design allows us to add support for new chains without modifying the three core modules.
We explain the functions of each module in Section~\ref{sec:endpoint}.

The \textbf{Oracle} is an third party service that provides a mechanism to, independently of the other \System{} components, read 
a block header from one chain and send it to another chain. In theory, this Oracle can be any third party service that provides this mechanism, but in practice
we expect to use Chainlink~\cite{chainlink17,chainlink21}, which is the current industry leader for decentralized oracle networks.

The \textbf{Relayer} is an off-chain service that is similar in function to an Oracle,
but instead of fetching block headers it fetches the proof for a specified transaction.

To ensure valid delivery, the only requirement is that for any given message sent using the \System{} protocol,
the Oracle and Relayer must be independent of each other.
The protocol itself does not require any specific implementation of a Relayer, and in theory the users of \System{} could even implement their own 
Relayer service. This design allows users to be sure that the Relayer cannot collude with the Oracle, and this independence
is what allows us to implement trustless validated delivery, as shown in Section~\ref{sec:tvd}.
In practice, \System{} provides the Relayer service while the Oracle is handled by Chainlink's decentralized oracle network and associated
consensus mechanisms.

\subsection{\System{} protocol}
\label{sec:protocol}

Figure~\ref{fig:commflow} illustrates the steps involved in the valid delivery of a single \System{} message.
Each encircled number in the figure represents a step of the protocol and corresponds to a paragraph in this section.
This section walks through the example of a user application on Chain $A$ sending a single message
to a user application on Chain $B$ via \System{}.
In Section~\ref{sec:casestudy}, we describe how the various components and protocol steps are implemented in the case
of sending messages between two Ethereum Virtual Vachines.

\textbf{Step 1:} The user application on chain $A$ (App $A$) executes some series of actions as part of transaction $T$. We uniquely
identify transaction $T$ by the transaction identifier \texttt{t}---the format of this identifier may vary depending on the type of chain $A$.
A step included in transaction $T$ is the transmission of a message over \System{} with valid delivery conditioned on $T$.
For illustration purposes, and without loss of generality, we assume that in this scenario App $A$ is using our reference Relayer.
App $A$ sends a request to the \System{} Communicator containing the following information:
\begin{itemize}
    \item \texttt{t}: The unique transaction identifier for $T$.
    \item \texttt{dst}: A global identifier pointing to a smart contract on chain $B$.
    \item \texttt{payload}: Any data that App $A$ wishes to send to App $B$.
    \item \texttt{relayer\_args}: Arguments describing payment information in the event that App $A$ wishes to use the reference Relayer.
\end{itemize}

\textbf{Step 2:} The Communicator constructs a \System{} packet containing \texttt{dst} and \texttt{payload}, referred to as \texttt{Packet(dst, payload)}, and sends it, along with \texttt{t} and \texttt{relayer\_args}, to the Validator.

\textbf{Step 3:} The Validator sends \texttt{t} and \texttt{dst} to Network. This step notifies Network that the block header for the current block on chain $A$ needs to be sent to chain $B$.

\textbf{Step 4:} Validator forwards \texttt{Packet(dst, payload)}, \texttt{t}, and \texttt{relayer\_args} to the Relayer, notifying the Relayer that the transaction proof for $T$ needs to be prefetched and eventually sent to chain $B$. This happens concurrently with \textbf{Step 3}.

\textbf{Step 5:} Network sends \texttt{dst} and the block ID of the current transaction (\texttt{cur\_blk\_id}) to the Oracle. This notifies the Oracle
to fetch the block header for the current block on chain $A$ and send it to chain $B$. In the event that multiple \System{} transactions occurred in the same block, \textbf{Step 5} is only executed once.

\textbf{Step 6:} Oracle reads the block header (\texttt{blk\_hdr}) from chain $A$.

\textbf{Step 7:} The Relayer reads the transaction proof associated with transaction $T$ (\texttt{proof(t)}) from chain $A$,
and stores if off-chain. \textbf{Steps 6} and \textbf{7} occur asynchronously to each other.

\textbf{Step 8:} The Oracle confirms that the block corresponding to \texttt{blk\_hdr} is stably committed on chain $A$ and then sends \texttt{blk\_hdr} to Network on chain $B$. The mechanism for determining
when this happens varies per chain, but will typically involve waiting for some number of block confirmations.

\textbf{Step 9:} Network sends the block hash, specified as \texttt{blk\_hdr\_hash}, to the Validator.

\textbf{Step 10:} The Validator forwards \texttt{blk\_hdr\_hash} to the Relayer.

\textbf{Step 11:} After receiving \texttt{blk\_hdr\_hash}, the Relayer sends a list of any \texttt{Packet(dst, payload), t, proof(t)} tuples that match
the current block. In the event that multiple users simultaneously send messages between the same endpoints, there may be multiple packets and associated transaction proofs within the same block.

\textbf{Step 12:} The Validator uses the received transaction proofs in conjunction with the block headers stored by Network to validate whether the associated
transaction $T$ is valid and committed. If the block header and transaction proof do not match, then the message is discarded. If they do match, then
\texttt{Packet(dst, payload)} is sent to the Communicator.

\textbf{Step 13:} The Communicator emits \texttt{Packet(dst, payload)} to App $B$.

\subsection{Achieving trustless valid delivery}
\label{sec:tvd}

\textbf{Trustlessness:} At the crux of \System{}'s design is the idea that the user need not trust the components of \System{}.
Instead of requiring trust, which is a strong condition, we only require the weaker condition of \emph{independence} between the Oracle and Relayer.
This requirement of independence instead of trust is one aspect of what allows \System{} to be efficient and lightweight.
As long as there is no malicious collusion between the Oracle and Relayer, then \System{} guarantees valid delivery.

\textbf{Valid delivery:} By the \System{} protocol shown in Section~\ref{sec:protocol}, a message $m$ is delivered by the Communicator to the user application if and only if the transaction proof
for the transaction $t$ associated with $m$ can be validated in \textbf{Step 12}. This validation step will succeed if and only if the block header and the
transaction proof match, which will only occur in the following two scenarios:

\begin{enumerate}
    \item The block header provided by the Oracle and the transaction proof provided by the Relayer are both valid.
    \item The block header provided by the Oracle and the transaction proof provided by the Relayer are both invalid, but still match.
\end{enumerate}

Scenario 2 can only happen if the Oracle and the Relayer collude, as it is statistically impossible to send a transaction proof that
can be validated against a block header without knowledge of that specific block header, and vice versa.
However, \System{}'s design eliminates the possibility of collusion, as outlined in Section~\ref{sec:intro}.
Thus, if a message is delivered to the user application on the receiver side, it is guaranteed to meet the properties of valid delivery.

As outlined in Section~\ref{sec:valid-delivery}, a communication protocol which can guarantee
\emph{trustless} valid delivery, namely valid delivery without placing trust in intermediary entities or tokens, is the ideal solution to enable cross-chain transactions.
\System{} is the first and only system to proved this property. This fact will drive user adoption of \System{} as the preferred method of cross-chain messaging.

\subsection{\SystemEndpoint{}}
\label{sec:endpoint}

A \SystemEndpoint{} is currently implemented as a series of smart contracts on each chain included in the \System{} network.
The core functionality of a \SystemEndpoint{} is encapsulated in three modules: the Communication, Validation, and Network.
These modules act in a manner similar to a network stack, with messages sent down the stack on the
sender side---Communicator to Validator to Network---then up the stack on the recipient side.

In addition to the core modules, \SystemEndpoint{} can be extended via Libraries, which are auxiliary smart contracts that define how communication
for a specific chain should be handled. Each chain in the \System{} network has an associated Library, and each Endpoint includes a copy
of every Library. This modular design allows the \System{} network to be quickly and easily extended to include new chains on demand. In addition,
communication between two chains only requires that their respective libraries be present on both ends, making \System{} a fully-connected network
with the ability to orchestrate transactions between any pair of nodes.

\subsection{\SystemEndpoint{} cost scalability}
\label{sec:endpoint-scalability}

As many readers will likely point out, running smart contracts on Layer 1 chains can be cost prohibitive,
especially as the amount of stored data increases.
To make the \SystemEndpoint{} practical, it was necessary for us to design the most lightweight client possible.
Previous work on trustless cross-chain validation through cross-chain state machine replication (SMR), such as
Golden Gate~\cite{goldengate}, could cost millions of dollars per day to run on popular Layer 1 chains like Ethereum.

To solve this problem, we set out to design the most lightweight client possible. 
Our key observation is that replicating and storing block headers within the
client is not necessary. Rather, we delegate the task of fetching the necessary cross-chain headers
and transaction proofs to off-chain entities: the Oracle and Relayer.
This results in \SystemEndpoint{}s being incredibly lightweight, making them cost-effective even on
notoriously expensive~\cite{spain_et_al:OASIcs:2020:11973} chains like Ethereum.

\section{Case Study: \System{} on EVM}
\label{sec:casestudy}

\begin{figure}
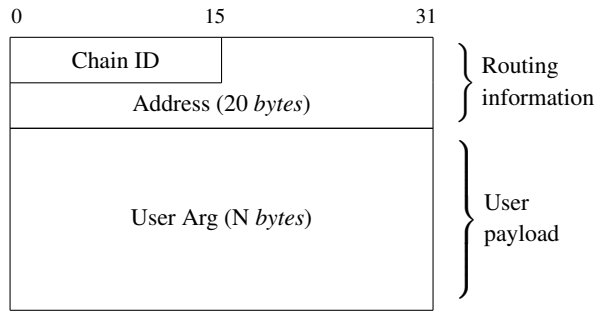

    \begin{bytefield}[bitformatting={\footnotesize},
                      bitwidth=0.5em,
                      boxformatting={\small\centering}]{32}
        \bitheader{0,15,31} \\
        \begin{rightwordgroup}{\small{Routing} \\ \small{information}}
            \bitbox{16}{Chain ID} \bitbox[tr]{16}{} \\
            \wordbox[lrb]{1}{Address (20 \textit{bytes})}
        \end{rightwordgroup} \\
        \begin{rightwordgroup}{\small{User} \\ \small{payload}}
            \wordbox[tlrb]{4}{User Arg (N \textit{bytes})} 
        \end{rightwordgroup}
    \end{bytefield}
        
    \caption{\System{} packet layout for EVM endpoints.}
    \label{fig:evm-packet}
\end{figure}

In this section, we briefly describe the details of how we implemented support for running
\System{} on Ethereum Virtual Machines (EVMs)~\cite{hildenbrandt2018kevm}. For the sake of brevity, we focus on
the aspects of the system whose implementation is likely to vary by chain and highlight
how our implementation handles the specific requirements of the Ethereum chain.
As mentioned in Section~\ref{sec:components}, the current version of \System{} relies
on Chainlink to provide the Oracle service, and expects users to use the Relayer service that we provide.

\textbf{\System{} packet:} The format of the \System{} packet will vary depending on the source and
destination chains. We illustrate the precise layout of the \System{} packet for EVM endpoints~\cite{solidity} in
Figure~\ref{fig:evm-packet}. Each field functions as follows:

\vspace{1mm}

\begin{center}
\begin{tabular}{l|l}
     \textit{Field} & \textit{Description} \\
     \hline
     Chain ID & A unique identifier for each \\
              & chain on the \System{} network.\\
     \hline
     Address & The address of the recipient smart\\
             & contract on the destination chain.\\
    \hline
    User Arg & The payload sent by the \\
    0 -- N   & user application---in EVM this\\
             & can contain up to N-byte argument.\\
\end{tabular}
\end{center}

\vspace{2mm}

\textbf{Sender-side chain transaction stability:} To ensure that the
message transaction is stable on the source chain, we rely on the inherent properties
of decentralized oracle networks---the Oracle will only notify the destination chain
of a particular block header after it hears some number of block confirmations, which
in the case of Ethereum is 15. Precisely speaking, \textbf{Step 8} of the \System{} protocol 
(Section~\ref{sec:protocol}) will only execute after the Oracle hears 15 block confirmations on chain $A$.

\textbf{\SystemEndpoint{}:} We implement the \SystemEndpoint{} as a series of smart contracts, composed of
the four main modules we describe in Section~\ref{sec:endpoint}. For most existing blockchains,
including the Ethereum blockchain, we are able to implement the Communicator, Validator, and Network each as
separate smart contracts. However, this design does not preclude the implementation
of \SystemEndpoint{} on (future) chains with different requirements.

The Library component of the \SystemEndpoint{} is the key to providing support for the Ethereum blockchain
in this case study. We implement a Library to handle the construction of the EVM-specific \System{} packet
shown in Figure~\ref{fig:evm-packet} and handle the encoding and decoding of EVM smart contract address information.

An additional responsibility of the Library is to handle the actual computation involved
in validating the transaction proof. Our EVM Library handles Merkle-Patricia Tree
validation~\cite{lu2019blockchain} for transactions on an EVM block, which we base on an open-source
implementation by Golden Gate~\cite{goldengate}.

\section{Applications on \System{}}
\label{sec:apps}

\textbf{Cross-chain decentralized exchange:} As briefly described in Section~\ref{sec:exchanges}, \System{} enables a cross-chain DEX (cross-chain bridge)
that deals exclusively in native assets. Contrary to existing DEX designs that issue wrapped tokens or go through intermediary sidechains,
a DEX built using \System{} to send messages between chains can be built such that liquidity pools exist on both chains, and users can simply
deposit their native assets in one pool and withdraw native assets from another. \System{}'s messaging primitive is powerful enough to enable
direct bridges (1:1 pricing), automated market making ($ab=k$ pricing), and any other derivation (such as one similar to Curve DAO pricing~\cite{warren20170x}).
The guarantee of valid delivery that \System{} provides enables a wide range of decentralized exchange applications.

\textbf{Multi-chain yield aggregator:} Current yield aggregators typically operate within the confines of single chain ecosystems,
with projects such as Yearn Finance~\cite{yearnfinance} enabling yield aggregation using single chain strategies.
One key weakness of these single chain yield aggregation systems it that they cannot take advantage of
any yield opportunities outside of their current ecosystem, potentially missing out on many of the best yields. A yield aggregator that uses \System{} for cross-chain
transactions would allow for strategies that tap into the best opportunities across all ecosystems, increasing access to high yield opportunities
and enabling users to take advantage of market inefficiency. A multi-chain yield aggregator would be strictly better than a single-chain yield aggregator, as
in the worst case the strategy would degrade to taking advantages of opportunities on only one chain, and in the best case it would have exponentially more
opportunities to choose from.

\textbf{Multi-chain lending:} Today, users have no easy way to take advantage of opportunities on chains where they do not hold assets.
For example, suppose that a user with assets consolidated in ETH wants to take advantage of an opportunity on Polygon~\cite{polygon}. Their
choices are to either (1) move their entire asset base to another chain and convert it to the desired currency, or (2) lend their assets on Ethereum, borrow the
desired asset, and then bridge that asset to the destination chain. \System{} enables a lending protocol that would allow the user to keep their entire
asset base in-place on Ethereum, lend it out, then borrow directly in MATIC on Polygon. This eliminates intermediary costs such as bridge and swap fees.

These three examples represent but a tiny fraction of the many possibilities that \System{} enables.
By leveraging \System{}, developers will be able to write their applications without worrying
about differing semantics between inter- and intra-chain transactions, and users will be able to freely move
liquidity across chains. We look forward to the creative new applications that the community will develop given the power of
trustless cross-chain transactions.

\section{Conclusion}
\label{sec:conc}

This paper introduced the design and implementation of \System{}, the first
trustless omnichain interoperability platform that does not involve any intermediate transactions.
We showed that by leveraging two independent, untrusted off-chain entities, the Oracle and Relayer,
\System{} is able to achieve valid delivery without requiring costly cross-chain state machine replication
or intermediary tokens. Our protocol is designed in a way that does not preclude the use of arbitrary relayer services,
which ensures that there is no collusion between the Relayer and Oracle.
The \System{} protocol enables native transactions between supported chains,
while the novel \SystemEndpoint{} design can be easily extended to support any chain. In addition to this,
our Endpoint design is lightweight enough to run on expensive Layer 1 chains such as Ethereum without
incurring prohibitive costs.
We presented a case study of how to implement support for EVM-based chains in \System{}, using a reference
Relayer implementation in conjunction with Chainlink's decentralized oracle network to enable cross-chain
transactions through \System{}.

\System{} is the backbone that will connect the various disjoint blockchain ecosystems and
allow frictionless movement of liquidity, data, and ideas between chains and communities.

{\footnotesize
\bibliographystyle{acm}
\bibliography{lz-bib}}

\end{document}